\input phyzzx
\hoffset=0.375in
\overfullrule=0pt

\def\max{{\rm max}}

\twelvepoint
\font\bigfont=cmr17
\centerline{\bigfont Pixel Lensing Search For Bright Microlensing Events}
\smallskip
\centerline{\bigfont and Variables in the Galactic Bulge}
\bigskip
\centerline{{\bf Andrew Gould}\footnote{1}{Alfred P.\ Sloan Foundation Fellow}
and {\bf D.\ L.\ DePoy}}
\smallskip
\centerline{Dept of Astronomy, Ohio State University, Columbus, OH 43210}
\smallskip
\centerline{gould,depoy@astronomy.ohio-state.edu}
\bigskip
\centerline{\bf Abstract}
\singlespace 
%\doublespace

	We describe a new method to search for gravitational microlensing 
toward the Galactic bulge that employs a small camera rather than a 
conventional telescope and probes new regions of parameter space.  
The small aperture ($\sim65\,$mm) permits detection of stellar flux variations
corresponding to magnitudes $7\lsim I \lsim 16$, whereas current searches are 
restricted by saturation to $I\gsim 15$.  The large pixel size 
($\sim10''$) and $\sim(6\,\rm deg)^2$ field of view allows  
observation of
the entire bulge with a few pointings.  With this large pixel size (and with 
the even larger $30''$ PSF that we advocate) most bulge stars are unresolved, 
so one is in the regime of pixel lensing:  microlensing and other forms of 
stellar variation are detected from the difference of pixel counts in
successive images.  We 
identify three principal uses of such a search.  First, the observations are 
analogous to normal pixel lensing observations of the bulge of M31, but are 
carried out under conditions where the detected events can be followed up in 
detail.  This permits crucial checks on the systematics of the M31 searches.  
Second, the search gives a complete inventory of bright bulge variables.  
Third, ``extreme microlensing events'' (EMEs) can be found in real time.  
EMEs are
events with maximum magnifications $A_\max\sim200$ which, if they were 
observed intensively from two observatories, could yield the mass, distance, 
and speed of the gravitational
lens.  The instrumentation required to carry out the observations
is inexpensive.  The observations could be made in parallel with 
existing microlensing searches and/or follow-up observations.  The data
reduction is much simpler than in ordinary pixel lensing because the PSF can be
fixed by the optics and so does not vary with atmospheric conditions.

\bigskip
Subject Headings: gravitational lensing -- stars: variables: other
\smallskip
\centerline{submitted to {\it The Astrophysical Journal}: July 4, 1997}
\smallskip
\centerline{Preprint: OSU-TA-17/97}

\endpage
%\normalspace
\chapter{Introduction}

	Pixel lensing is the search for microlensing of unresolved stars by
subtracting successive images of the same field.  Crotts (1992) and Baillon 
et al.\ (1993) suggested a method to probe the halo
of M31 for massive compact halo objects (MACHOs).  In fact, it is the only
possible method to conduct such a search from the ground because very few
M31 stars are resolved.  By contrast, microlensing searches in nearer star
fields like the Large Magellanic Cloud (LMC) (Alcock et al.\ 1997b MACHO; 
Ansari et al.\ 1996 EROS) and the Galactic bulge (Udalski et al.\ 1994 OGLE; 
Alcock et al.\ 1997a MACHO; Alard 1997 DUO) have used the classical technique 
suggested by Paczy\'nski (1986): they perform repeat photometry on 
individually identified 
stars.  Since these fields contain millions of resolved stars, the classical
approach works quite well and has produced important results.

	Nevertheless, one could in principle apply pixel lensing techniques
to the LMC and Galactic bulge.  Melchior (1995) has carried out a
pilot pixel lensing study of the LMC using archival EROS data, and
Tomaney (1997, private communication) has begun pixel lensing searches of
both the bulge and the LMC using archival MACHO data.  Heretofore, however,
the primary motivation for pixel lensing searches in these resolved star fields
has been to make use of the additional sources that lie below the detection 
threshold, and so to increase the sensitivity of the existing experiments.

	Here we propose a radically different type of pixel lensing experiment:
use a small camera with large pixels and with deliberately degraded optics to
search for microlensing and other forms of stellar variation toward the 
Galactic bulge.  The initial motivation for this proposal is to mimic 
(and thereby to better understand) pixel lensing searches toward M31.  However,
such a search would yield important information on the characteristics of
the bulge as well.

\chapter{M31 Pixel Lensing}

	When Crotts (1992 Columbia-VATT) and Baillon et al.\ (1993 AGAPE) 
first proposed pixel
lensing of M31, the idea was greeted with extreme skepticism even in the
microlensing community.  The principal worry was that it would be impossible
to take proper account of seeing variation, so the image differences would
be too noisy to detect stellar variations.  This objection is now clearly
laid to rest by the beautiful difference images of Tomaney \& Crotts (1997)
which show hundreds of variable stars, many with variations at the level
of a few percent of the background galaxy surface brightness.  AGAPE also
has good sensitivity to variable stars although their technique does not
yield such striking visual representations (Ansari et al.\ 1997).  
	Nevertheless, neither group has reported a confirmed microlensing
event.  Part of the reason is a shortage of telescope time, itself
partly engendered by the initial skepticism.  There are a number of 
microlensing candidates (Crotts \& Tomaney 1997), but the baselines are not
long enough to rule out the possibility that these are long period variables.  

	However, even when more data are acquired, confirmation of
events toward M31 will remain more difficult than toward the LMC and
the Galactic bulge.  There are two main reasons for this increased difficulty. 
First, the 
signal-to-noise (S/N) ratio is inevitably much lower for the more distant
sources in M31.  For LMC sources, and especially for Galactic bulge sources,
substantial additional confidence in the microlensing 
interpretation of the events comes
from the close fit of the light curves to the simple 
three-parameter Paczy\'nski (1986) form.  A second related problem is that 
only for the brightest M31 sources is the S/N sufficiently high to allow
precise measurements of the light curve, but these are just the sources most
likely to be intrinsically variable.  While most bright variables repeat
(and so can be identified given a long enough baseline) there is always the
possibility that a new rare class of variable will more closely mimic
microlensing.  Indeed, MACHO discovered a new such class of bright variable 
toward the LMC, called ``bumpers'' (Alcock et al.\ 1996a).  
These (as well as several other previously known classes of 
variables) are easily removed
from the LMC search catalog simply by excluding the bright stars.  
Moreover, bright stars in the LMC can always be followed up spectroscopically
if they raise any suspicions.  The situation is quite different toward M31.
Bright stars cannot be eliminated without also removing all the high S/N
events. Moreover, spectroscopic follow-up is very difficult and in many cases 
impossible.  Hence, the interpretation of M31 lensing candidates will always
be less secure than those found toward the LMC or the Galactic bulge.

	Both the disk and the bulge of M31 are being monitored for pixel
lensing events.  In certain respects, the M31 disk fields are similar to those
of the LMC and the M31 bulge fields are similar to those of the Milky Way 
bulge.  The M31 disk is like the LMC in that both have
significant populations of early type stars and in that both lines of
sight are relatively devoid of known populations of foreground stars.  
Thus, if more
than a trickle of events are detected, these are probably due to a
previously unknown population of halo objects.  For the LMC events, these would
mostly be in the Milky Way halo, whereas for the M31 disk events, they
would be in the M31 halo.  By contrast, lensing of Milky Way or M31 bulge 
stars can be caused by other bulge stars, so one is learning primarily about 
a known population.  Of course, one may learn unexpected things
about these ``known'' populations.  Indeed the ``known'' population of lenses
toward the Galactic bulge is yielding a number of surprises including a
higher-than-expected optical depth and a perplexing excess of short events
(Udalski et al.\ 1994; Alcock et al.\ 1997a; Han \& Gould 1996; Han 1997).

\chapter{Mimicking M31 Bulge Pixel Lensing Using the Galactic Bulge}

	Here we propose to mimic (and so learn more about) M31 bulge pixel 
lensing by artificially degrading the observing conditions toward the Galactic 
bulge.  To zeroth order, the source populations and the lens populations 
should be similar in the two bulges.  If they are not, this itself would 
be very interesting.  One might also think about mimicking M31 disk pixel
lensing by degrading the observing conditions toward the LMC.

\section{Observational Parameters}

	A moderately aggressive program for M31 bulge pixel lensing might 
obtain 1 hour exposures every clear night on a 1.3 m telescope with $1''$ 
seeing and $0.\hskip-2pt ''3$ pixels.
A $2048\times 2048$ CCD would then cover $100\,\rm arcmin^2$.  To precisely
mimic these conditions for observations of the Milky Way bulge (which is
100 times closer) one would want $30''$ pixels
and a $100''$ point spread function
(PSF).  A $2048\times 2048$ CCD would cover $300\, \rm deg^2$. For a camera 
with a 65 mm diameter primary optic, the exposure time should be
2 minutes.   More accurately, the differential extinction in the two directions
must be taken into account to maintain similar observing conditions.
For definiteness,
we focus on $I$ band. For M31, $A_I\sim 0.15$.  For the Milky Way bulge as a 
whole, the extinction is highly variable.  However, there are broad regions
of the southern bulge $(b\lsim -2^\circ)$ 
for which the extinction is moderately
low, $A_I\lsim 1.3$.  For convenience, we adopt $A_I=1.3$.  For
comparison, we note that 
the extinction in the ``Blanco A region'' of Baade's Window is
$A_I=0.83$ (Gould, Popowski, \& Terndrup 1997).  The difference, 
$\Delta A_I=1.15$, then implies that the Milky Way observations require
an exposure time longer by a factor 2.9, or 6 minutes.

\section{Optimal Parameters}
	
	While the above parameters would closely reproduce the M31 
observations, altering them slightly would give substantially more 
information from the bulge pixel lensing observations without compromising 
their value as tests of M31 pixel lensing.  In particular, by reducing
the pixel size and PSF each by a factor $\sim3$ to $10''$ and $30''$
respectively, the S/N of detected variations would increase by the same
factor.  In addition, one would reduce confusion between neighboring variable
sources.  Similarly, one could increase the exposure time (or number of 
exposures) to obtain better S/N.  To simulate M31 pixel lensing, one could
then simply convolve the images with a larger PSF and add appropriate noise.

	One might then ask: why not go to the limit of small pixels and large
apertures?  In particular, why not use the Galactic
bulge microlensing observations
themselves, and simply degrade these to the M31 pixel lensing conditions?
Unfortunately, the bulge microlensing observations cannot be degraded to mimic
M31 observations because they
saturate at $I\sim15$ which corresponds to $I\sim24$ in M31, well below the
threshold of detection.  In principle, one could use the bulge microlensing
telescopes and drastically reduce the exposure times to avoid saturation.
However, given the requirements of readout and pointing, this would 
take an inordinate amount of time at the expense of the basic microlensing
search.  A small camera is the most efficient way to cover a large field
of view while avoiding saturation.

	We are led to a choice of $10''$ pixels by the following 
considerations.  The camera could be mounted behind the secondary of a 
telescope
already dedicated to microlensing observations, either the searches themselves
or follow-up observations such as are now being carried out by 
PLANET (Albrow et al.\ 1996) and GMAN (Alcock et al.\ 1996b).  The observations
would then be carried out in parallel with the primary microlensing 
observations.  The field of view should then be large enough so that the
resulting
images overlap and cover most or all of the ($\sim50\,\rm deg^2$) of interest
in the southern bulge.  In particular, one expects that the follow-up 
observations will attempt to monitor ${\cal O}(20)$ microlensed sources
scattered over the southern bulge, so a $\sim6\,$deg field (corresponding
to $10''$ pixels and a $2048\times 2048$ detector) should ensure overlapping 
coverage of the entire region.

\section{Pixelization Noise}

	In order to form a smooth mosaic from many individual images, the
PSF must be substantially larger than a pixel.  This is also an important
consideration for reducing noise when taking the difference of successive
images.  Gould (1996) showed that the ratio of ``pixelization noise''
induced by finite pixel size to photon noise is given by
$${%F^2_
{{\rm pixel}\ {\rm noise}}\over %F^2_
{{\rm photon}\ {\rm noise}}} \sim
{\kappa n_*\over 500 \sigma^4},\eqn\pixelnoise$$
where $\sigma$ is the Gaussian width of the PSF in pixels, $\kappa$ is the
ratio of the galaxy surface brightness to the total surface brightness
(galaxy plus sky), and $n_*$ is the number of photons collected from a
``fluctuation magnitude star'' ($M_I\sim -1.2$) during an exposure.  Much
of the bulge is above sky ($\kappa\sim1$).  If the observations are carried
out in parallel with follow-up of giants, then the exposures will be 
1 or 2 minutes.  For definiteness, we assume 2 minutes.  A $30''$ PSF and
$10''$ pixels imply $\sigma\sim1.3$.  The ratio in equation \pixelnoise\
is then $\sim 0.5$.  Hence, if the PSF were substantially smaller, the
pixelization noise would dominate the photon noise.

\section{Signal-to-Noise Ratio}

	Sensitivity is defined slightly differently for pixel lensing 
experiments than it is for ordinary photometry.  In the latter, one measures
the total flux from a star $F$ and then converts to a magnitude
$I = -2.5\log F + C$ where $C$ is a constant.  In pixel lensing, one detects
only $\Delta F$, the difference in flux from a star between two epochs.  One
converts this to a magnitude using the same formula, $I = -2.5\log \Delta
F + C$.  Note that this ``difference magnitude'' is {\it not} the change
in the magnitude of the star (which is unknown).  Rather, it is the magnitude
of an imaginary star whose flux is the same as the difference of the fluxes
between epochs.

	Typical bulge fields (e.g.\ Baade's Window) have a dereddened surface 
brightness $I\sim 17.6$ mag arcsec$^{-2}$ (Terndrup 1988).  
We again assume a typical extinction
of $A_I\sim 1.3$.  In a two minute exposure by a 65 mm 
diameter camera with 25\% overall efficiency, 
a $10''$ pixel will collect $\sim1500$ electrons.  Hence,
commercially available CCDs with readout noise of $\sim15e^-$ will not
seriously degrade the S/N.

	The area of the PSF is $\Omega_{\rm psf}\sim 0.75\,\rm arcmin^2$,
implying that the (reddened) Galaxy background light in a PSF is equivalent
to $I\sim 10.3$, i.e., $\sim 400 e^{-}\,\rm s^{-1}$.
	If the observations are made in parallel with 
follow-up observations, one
may expect that the total exposure time at each point will be about 30 minutes
per night.
This means that a difference magnitude of $I=15$ would be detectable at the
$11\,\sigma$ level.  This is about 7 times better than the S/N for the same 
fluctuation taking place in M31 and observed for 1 hour with a 1.3 m 
telescope in $1''$ seeing.

	Assuming that the well depth of the pixels is $\sim 100,000\,e^-$,
saturation would occur for sources at $I\sim7$ in a two minute exposure. The
only bulge variable sources likely to exceed this limit are supernovae.

\section{Uniformity of PSF}

	In standard pixel lensing, the PSF varies with the atmospheric
conditions.  Correcting for these variations is challenging (Tomaney \& Crotts
1997; Ansari et al.\ 1997).  By contrast, the $30''$ PSF proposed here could 
be set by the optics
and would not vary in time.  The problems of forming mosaics and differencing
successive images would therefore be substantially reduced.  However, in order
to take full advantage of this simplification, the PSF must also 
be uniform over
the image.  Otherwise, the PSF would not actually be the same on images
whose centers are offset from one another.

\section{Optics}

	A uniform PSF can be achieved simply by defocusing an appropriate 
optical design.  Experience with reflecting telescopes makes this assertion 
seem counter-intuitive because an out-of-focus telescope has a donut-shaped
PSF.  However, this shape is a caused by the fact that the secondary occults
the central portion of the mirror, a problem that does not affect lens-based
cameras.  

	For a CCD with 15 $\mu$m pixels, the focal length of the system should 
be roughly 310 mm ($f/4.75$ for a 65 mm entrance aperture).  
The simplest choice
for the optical design would appear to be an achromatic doublet.  However, 
this has off-axis aberrations that are too severe
to provide a uniform PSF.  That is, by the 
time the on-axis image is made sufficiently large by defocussing the lens, the
off-axis images are very unsymmetric and substantially larger than the on-axis
image.   A classic Cooke Triplett (see Smith 1990 or Kingslake 1965) provides 
one good solution.  When in focus, it gives excellent ($<1$ pixel) PSFs 
over the 
entire field covered by a $2048\times 2048$ detector array.  Paradoxically, 
we do not want such small PSFs.  However, the residual aberrations are small 
enough that even after modest defocussing the on-axis and off-axis images
are very similar.  Other solutions are also possible.

\chapter{Science}

\section{Comparison with M31 Pixel Lensing}

	The 65 mm pixel lensing data can be used to perform several types of
checks on M31 pixel lensing.  The data can be degraded to reproduce the
characteristics of a series of M31 images by convolving them with a larger
PSF and adding noise.  They can then be searched for
microlensing events using exactly the same algorithms that are used for M31.
The resulting lens candidates can be checked in several different ways.
First one could examine the undegraded 65 mm pixel lensing data which, from
\S\ 3.4, should have 7 times better S/N.  If the event is detectable at all
from the degraded data, it should be very easy to tell whether it is indeed
microlensing from the undegraded data.  Second, much of the bulge will be
covered in much finer detail by the regular microlensing surveys.  While the
peak of typical pixel lensing events will not be visible in the regular
data set because of saturation,
there could well be good photometry of the source star away from the event
which would give important clues to the nature of the source.  Third,
from the 65 mm pixel lensing data alone, or perhaps in combination with the
regular microlensing survey data, the precise position of
the source could be found which would permit spectroscopic follow-up.

	One could also empirically evaluate the pixel lensing efficiency as
a function of peak flux variation and effective timescale.
After the algorithm had found all of the events that it thought were 
microlensing in the degraded image series, one could search both the
undegraded images and the regular microlensing survey catalogs for all 
events.  The latter are likely to be close to complete (or at least have
well understood completeness) for events that could plausibly be detected
from the degraded images.  This procedure will automatically take account
of the effects of variables, both variables mistaken for microlensing events
and confusion caused by variables near genuine microlensing events.

\section{Bright Bulge Variables}

	Pixel lensing observations of the bulge would easily find all variables
with difference magnitudes $I<15$.  In principle, one could imagine 
processing the time series to extract much fainter variables.
However, the regular microlensing observations are more efficient at finding
these fainter variables.  They are insensitive to bright variables because
of saturation.

	A complete catalog of bright variables would be extremely useful for
bulge studies.  Previous studies cover only patches of parameter space and
suffer from small statistics. Lloyd Evans (1976) carried out the first 
large-scale 
optical search for bright bulge variables.  He obtained 22 epochs for
 three fields, 
each about $0.33\,\rm deg^2$, at $(l,b)=
(1^\circ\hskip-2pt .4, -2^\circ\hskip-2pt .6)$,
$(0^\circ\hskip-2pt .9, -3^\circ\hskip-2pt .9)$, and
$(4^\circ\hskip-2pt .2, -5^\circ\hskip-2pt .1)$.  He found 57, 38, and 12
Miras, respectively.  
A subset of these stars were subsequently studied by 
Glass \& Feast
(1982) in the infrared where most of their luminosity is generated.
Terzan, Wehlau, \& Wehlau (1986) found 36 variables in $\sim 0.8\,\rm deg^2$
at $(l,b)=
(3^\circ\hskip-2pt .5, 4^\circ\hskip-2pt .0)$ observed at 57 epochs.
Whitelock, Feast, \& Catchpole (1991) made 10--12 $JHKL$ observations of 141
bulge sources at $7^\circ <|b|<8^\circ$ selected on the basis of the their
{\it IRAS} colors and found 113 to be Miras.  Weinberg (1992) studied 3170 
sources classified as variable by {\it IRAS} and further selected by
{\it IRAS} colors.  He concluded that these stars, thought to be mostly AGB
stars, lay in a highly barred distribution
although the effective magnitude limit of the sample allowed him to map
out only the near side of the bar.

	A complete sample of bright bulge variables would permit study of the
following questions.  First, what is the three dimensional distribution
of Miras?  If the bar found by Weinberg (1992) is confirmed, how does it
relate to the bars found in the COBE light (Binney, Gerhard, \&
Spergel 1997) and in clump giants (Stanek et al.\ 1997)?  These two bars
do not agree with each other and neither agrees with the distribution of
RR Lyrae stars which show little if any detectable bar 
(D.\ Minniti 1997, private communication).  What is the relative number
of Miras and clump giants as a function of position in the bulge (and hence
of metallicity).  This should be a good indicator of the relative lifetimes
of these two phases of evolution as a function of metallicity.  How does
the period-luminosity relation vary with position (and hence metallicity)?

	While Miras are probably the largest class of bright variables,
others are also potentially interesting.
Are there Type II Cepheids in the
bulge?  To our knowledge, none have been discovered.  Type II Cepheids are
usually associated with metal-poor populations, so perhaps none should be
expected.  However, the $r^{-3.5}$ profile of RR Lyrae stars (a metal-poor
tracer) continues in toward the Galactic center to at least 500 pc 
(D.\ Minniti 1997, private communication), so the metal poor population
may be quite dense close to the center.  In addition, many novae will be
discovered and there may be other rarer classes of variables as well.

\section{Extreme Microlensing Events}

	Gould (1997) pointed out that extreme microlensing events (EMEs)
could potentially be used to measure individual masses, distances, and 
transverse velocities for up to 30 bulge lensing events per year.
EMEs are lensing events of main-sequence stars with maximum magnification
$A_\max\sim 200$.  Their small impact parameters $\beta\sim 1/200$ imply
that it is often possible to measure both the ``parallax'' and the ``proper
motion'' of an EME.  The parallax measurement requires simultaneous
follow-up observations from two sites separated by about one Earth radius, and
yields the size of the Einstein ring
projected onto the observer plane, $\tilde r_e$.  The proper motion measurement
yields the angular Einstein radius, $\theta_e$.  These two quantities can be 
combined with
the measured timescale of the event to extract the three physical parameters.
For example, the mass is given by $M = (c^2/4 G)\tilde r_e\theta_e$.  

	The accuracy of the photometry needed to measure the parallax of an
EME is of order 1\% or better which is moderately challenging.  However,
the most difficult problem is to identify the event before its peak which is
a prerequisite for making intensive follow-up observations from two sites.
In principle, it is possible to make this identification by comparing two sets
of regular bulge microlensing search observations 
taken on successive nights.  Unfortunately,
the EME source stars are generally not in the template, so the EME cannot be
found by standard techniques.  A pixel lensing search is required.  
The regular bulge microlensing data and the 65 mm diameter camera data can
play complementary roles in such a search.

	 Typical events with magnification $A_\max\sim 200$ peak at 
$I_0\sim 13.2$.  Assuming an extinction $A_I\sim 1.3$ and an Einstein 
crossing time $t_e\sim 10\,$days, then 1 day before peak the star will be 
$I\sim 17.7$.
At this brightness, the ``new star'' would be clearly visible in
the regular microlensing search images, provided one knew where to look.
The main problem is how to process the $2\times 10^9$ pixels (covering
$\sim50\,\rm deg^2$) to find the new star.  Perhaps the simplest approach would
be to
degrade the PSF to $\sim5''$ so that, as with the 65 mm observations, the PSF 
would be independent of atmospheric conditions.  Then one coud directly 
subtract successive images with minimal processing.  Assuming 1 minute 
exposures on a 1 m telescope,
an $I\sim 17.7$ difference star would be detectable at the $16\,\sigma$ level.
The whole bulge will probably not be observed every night, but
even two days before peak, the difference star would be detectable at the
$8\,\sigma$ level.  Unfortunately, for every 40 such events that are
alerted on, only one will become an EME two days later.  The 65 mm observations
can play a complementary role of singling out those candidates that are
becoming EMEs. At 4 hours before peak the new star will be $I\sim 15.8$,  
which, according to \S\ 3.4, would be detectable at the $5.5\,\sigma$ level.
Even for a quarter night, it would be detectable at the $3\,\sigma$ level.
While this is far too low a significance to {\it choose} candidates, it is 
adequate to choose {\it among} the candidates found from the regular survey
data taken a night or two previously.  These secondary alerts would be rare
enough (less than one per night) so that they could be checked directly
using the follow-up telescope.

	Note that to maximize the effectiveness of this method of finding
EMEs, there should be 3 such cameras in operation, one in South America, 
one in Africa, and one in Australia.

{\bf Acknowledgements}:   We thank Tim Axelrod for making the suggestion
to use a small camera rather than degrading telescope observations to
do bulge pixel lensing.  AG thanks the Aspen Center for Physics for its
hospitality and for providing a most pleasant environment in which some
of the work was carried out.  Work by AG was supported in part by grant 
AST 94-20746 from the NSF and in part by grant NAG 5-3111 from NASA.
Work by DLD was supported in part by grant AST 95-30619 from the NSF.

\endpage
\Ref\alard{Alard, C.\ 1997, A\&A, 321, 424}
\Ref\albrow{Albrow, M., et al.\ 1996, in IAU Symp.\ 173 ed.\ C.\ S.\ 
Kochanek \& J.\ N.\ Hewitt) (Dordrecht: Kluwer), 227}
\Ref\Alcc{Alcock, C., et al.\ 1996a, ApJ, 461, 84}
\Ref\Alcc{Alcock, C., et al.\ 1996b, ApJ, 463, L67}
\Ref\Alca{Alcock, C., et al.\ 1997a, ApJ, 479, 119}
\Ref\Alcb{Alcock, C., et al.\ 1997b, ApJ, in press}
\Ref\Ansari{Ansari, R., et al.\ 1996, A\&A, 314, 94}
\Ref\Ansari{Ansari, R., et al.\ 1997, A\&A, in press}
\Ref\Aubourg{Aubourg, E., et al.\ 1993, Nature, 365, 623}
\Ref\bail{Baillon, P., Bouquet, A., Giraud-H\'eraud, Y., \& Kaplan, J.\ 1993,
A\&A, 277, 1}
\Ref\bgs{(Binney, J., Gerhard, O., \& Spergel, D.\ 1997, MNRAS, submitted}
\Ref\crotts{Crotts, A.\ P.\ S.\ 1992, ApJ, 399, L43}
\Ref\crotom{Crotts, A.\ P.\ S., \& Tomaney, A.\ 1996, ApJ, 473, L87}
\Ref\gf{Glass, I.\ S., \& Feast, M.\ W.\ 1982, MNRAS, 198, 199}
\Ref\gtwo{Gould, A.\ 1996, ApJ, 470, 201}
\Ref\gthree{Gould, A.\ 1997, ApJ, 480, 188}
\Ref\gpt{Gould, A., Popowski, P.,\& Terndrup, D.\ T.\ 1997, ApJ, submitted}
\Ref\hg{Han, C., \& Gould, A.\ 1997, ApJ, 484, 000} 
\Ref\hg{Han, C., \& Gould, A.\ 1996a, ApJ, 467, 540} %mass spectrum from times
\Ref\king{Kingslake, R.\ 1965, Applied Optics and Optical Engineering,
(New York: Academic Press)} 
\Ref\le{Lloyd Evans, T.\ 1976, MNRAS, 174, 169}
\Ref\melc{Melchior, A.-L.\ 1995, thesis, Coll\`ege de France, Paris}
\Ref\Pac{Paczy\'nski, B.\ 1986, ApJ, 304, 1}
\Ref\smith{Smith, W.\ J.\ 1990, Modern Optical Engineering (New York: McGraw 
Hill)}
\Ref\stan{Stanek, K.\ Z., Udalski, A., Szyma\'nski, M., Kaluzny, J., Kubiak, 
M., Mateo, M., \& Krzemin\'nski, W.\ ApJ, 477, 163}
\Ref\gpt{Terndrup, D.\ T.\ 1988, AJ, 96, 884}
\Ref\tww{Terzan, A., Wehlau, A.\ F., \& Wehlau, W.\ H.\ 1986, AJ, 92, 809}
\Ref\tomcro{Tomaney, A., \& Crotts, A.\ P.\ S.\ 1996, AJ, 112, 2872}
\Ref\oglea{Udalski, A., et al.\ %Szyma\'nski, M.,  
1994, Acta Astronomica 44, 165}% OGLE experiment
\Ref\weinberg{Weinberg, M.\ D.\ 1992, ApJ, 384, 81}
\Ref\wfc{Whitelock, P., Feast, M., \& Catchpole, R.\ 1991, MNRAS, 248, 276}

\refout
\endpage
%\figout
%\endpage
\bye